\documentclass[aps,prb,twocolumn,showpacs,floatfix]{revtex4}

\usepackage{graphicx}%

\begin{document}

\title{A Defective Graphene Phase Predicted to be a Room Temperature Ferromagnetic
Semiconductor}

\author{L. Pisani$^1$,  B. Montanari$^2$, and N. M. Harrison$^{1,3}$}

\affiliation{$^1$Department of Chemistry, Imperial College London, South Kensington campus, London SW7
 2AZ, United Kingdom \\ $^2$CCLRC Rutherford Appleton Laboratory, Chilton, Didcot, Oxfordshire OX11 0Q
X, United Kingdom \\ $^4$CCLRC Daresbury Laboratory, Daresbury, Warrington WA4 4AD, United Kingdom}

\pacs{pacs}
\date{\today}

\begin{abstract}
Theoretical calculations, based on hybrid exchange density functional theory,
are used to show that in graphene a periodic array of defects 
generates a ferromagnetic ground state at room temperature 
for unexpectedly large defect separations. 
This is demonstrated for defects that consist of a carbon vacancy in
which two of the dangling bonds are saturated with H atoms.
The magnetic coupling mechanism is analysed and found to be due to 
an instability in the $\pi$ electron system with respect to a long-range
spin polarisation characterised by alternation in the spin direction between 
adjacent carbon atoms. The disruption of the $\pi$-bonding opens a
semiconducting gap at the Fermi edge. The size of the energy gap and the 
magnetic coupling strength are strong functions of the defect separation and
can thus be controlled by varying the defect concentration.
The position of the semiconducting energy gap and the electron effective mass
are strongly spin-dependent and this is expected to result in a spin asymmetry 
in the transport properties of the system. 
A defective graphene sheet is therefore
a very promising material with an in-built mechanism for tailoring 
the properties of the spintronic devices of the future.
\end{abstract}


\maketitle

\section{Introduction}
\label{intro}
The miniaturisation of silicon devices has delivered exponentially 
growing performance for the past fifty years but quantum limits are now being 
reached as transistor gate lengths approach 5nm. Spintronics,
which exploits the electron spin rather 
than the charge, will operate on a sub 1nm length scale and
thus facilitate another generation of devices. 

The achievement of many of the promises of spintronics depends
on the ready availability of a material that is a semiconductor, 
ferromagnetic at room temperature, and with a highly tunable band gap
and magnetic coupling.  The controlled synthesis of a robust room 
temperature, ferromagnetic semiconductor on a nanoscale has yet to
be achieved.
Recent reports of high temperature ferromagnetism in metal-free 
carbon materials have generated a great deal of
interest~\cite{4}. Proton-bombardment of graphite~\cite{5} gives rise
to a ferromagnetic phase which x-ray circular dichroism measurements~\cite{xmcd}
have shown to originate from carbon $sp$ electrons.
Pyrolytic treatment of organic compounds~\cite{6} also gives rise to a
ferromagnetic matetieal. In both cases, the magnetic phase appears to be a minority 
phase in a non-magnetic host. The detailed composition and structure of 
the minority phase have not yet been determined but the concentrations of 
magnetic impurities have been measured sufficiently accurately to suggest that
the magnetism is not the result of contamination\cite{7}.

The occurrence of high temperature ferromagnetism in purely sp-bonded materials 
is a major challenge to current theoretical understanding of magnetic 
interaction mechanisms. A chemical or structural defect in a carbon-based 
material leads to the creation of local magnetic moments due to the presence 
of under- or over-coordinated atoms.
The presence of local moments needs to be accompanied by strong long-range 
coupling between them for the ferromagnetic order to survive thermal fluctuations
at room temperature. In extended graphene sheets, local moment formation
at point defects is now well established~\cite{8}.
A significant step forward in the study of the magnetic coupling between 
local defects has been the demonstration that the presence of spin moments at 
the edges of graphene ribbons can lead to an instability of the $\pi$-electron 
system, providing a mechanism for long range antiferromagnetic coupling 
which is robust with respect to competing instabilities involving charge 
ordering and geometric distortions~\cite{9,10,11}. 
More recently, Yazyev and Helm~\cite{yaz07} have studied defective graphene 
using density functional theory. They find a ferromagnetic ground state
which they assume to be of the itinerant type due to the metallicity of their system.  
Vozmediano {\it et al.}~\cite{voz} studied the interaction between localised moments
originating from lattice defects at large distances and concluded
that the transition temperature to the ferromagnetic RKKY state is
far below room temperature. 
Dugaev {\it et al.}~\cite{dug06} predicted that if the spin-orbit
interaction is taken into account, an energy gap opens at the Fermi
level and the RKKY model is not applicable.
A possibile source of high temperature ferromagnetism in carbon systems 
has been ascribed by Edwards {\it et al.}~\cite{edw} to  a limited screening of the 
interactions between itinerant electrons in impurity sp-bands. 
The expected suppression of the Stoner critical temperature 
(typical in the case of transition metal impurity d-bands) is considerably reduced
due to  a less effective screening of the electron interaction.  
In the present case we show a completely different scenario, 
i.e. that high temperature ferromagnetism in a carbon system
can be entirely supported by an insulating state of the electronic structure.
 
The current work demonstrates that local spin moments in defective graphene 
interact strongly over long distances and may therefore lead to a ferromagnetic semiconducting state 
at  room temperature.
The strong, long-range magnetic coupling necessary for 
magnetic ordering at room temperature in this semiconducting material 
is provided via a spin alternation mechanism that depends on the
presence of partially filled $\pi$ orbitals in the sp$_2$-bonded, 
bipartite lattice of graphene. Moreover, it is shown that the 
spin-dependent band gap and magnetic coupling strength can be varied 
through control of the defect concentration, opening up the possibility 
of highly tunable graphene-based spintronics nanodevices.
This study, therefore, reinforces the case, which has been growing
stronger and stronger since the ground-breaking practical realisation of isolated
graphene sheets~\cite{2}, for considering graphene
amongst the most promising materials for building the electronic
and spintronic devices of the future.

\section{Computational Details}
\label{det}
Spin localisation at defects and long range spin coupling are studied here using
 all electron, two-dimensionally periodic, hybrid exchange density functional 
theory (B3LYP~\cite{12,13,14}) which significantly extends the
reliability of the widely used
local and gradient corrected approximations to density functional theory in 
strongly interacting systems\cite{15,16,17}.
In the CRYSTAL package, used for this study, the crystalline wavefunctions
are expanded as a linear combination of atom centred Gaussian orbitals
(LCAO) with $s$, $p$, $d$, or $f$ symmetry. The calculations reported
here are all-electron, i.e., with no shape approximation to the ionic potential or
electron charge density. Basis sets of double valence quality (6-21G$^{*}$ for C 
and 6-31G$^{*}$ for H) are used.
A reciprocal space sampling on a Monkhorst-Pack grid of shrinking
factor equal to 6 is adopted after finding it
to be sufficient to converge the total energy to within $10^{-4}$ eV  per unit cell.

In the present work we consider three types of structure,
characterised by different arrangements of defects.
The first structure contains defects arranged in pairs
within a rectangular unit cell that is 25 \AA\ wide and of length 2$L$, which
is convenient for studying inter-defect coupling. 
The system under study is therefore infinitely periodic in the 2 dimensions parallel
to the graphene plane and non-periodic in the direction perpendicular to the plane. 
In Fig. 3 the unit cell containing a pair of defects is shown.
The defects are therefore arranged along parallel chains running along
the horizontal direction and the chains are 25 \AA\ apart.
The second structure is an hexagonal cell used to quantify inter-chain effects. 
The third structure is a triangular superlattice used for representing the band structure 
in a form compatible with that of graphene (Fig. 4).
All structures considered were initially relaxed using
a force-field model (GULP~\cite{17a}) in which the carbon-carbon and 
carbon-hydrogen interactions are described via the Brenner bond order potential~\cite{17a}.
In the case of the calculation of the stabilisation energy $\Delta$E (see below and Fig. 2),
full quantum mechanical relaxation was carried out 
for the structure with L=7.5\AA \ (see below)
and resulted in a small increase in $\Delta$E (see Fig. 2) of 25 meV.
As the additional relaxation is confined to the region of the defect, the
energy shift is expected to be depend weakly on the defect separation especially at 
large distances and thus 
full quantum mechanical relaxation was not required for larger structures.
The magnetic stablisation energy reported in Fig. 2, and the transition temperature based
on it, is therefore slightly underestimated.
In the analysis of the electronic band structure shown in Fig. 4,
where defects are displayed in a triangular lattice and are 20 \AA\ apart, 
we performed a full quantum-mechanical optimisation in order to quantify
the magnitude and decay law of the semiconducting gap.

\section{Results and Discussion} 
\label{res}
A number of point defects are possible. Here we choose to study a simple carbon
 vacancy in which two of the three dangling carbon bonds are saturated with 
H-atoms. This defect has been proposed previously, studied extensively~\cite{8}, 
and is displayed in Fig. 1. 
\vspace{.2cm}
\begin{figure}[!h]
\begin{center}
\includegraphics[scale=.75]{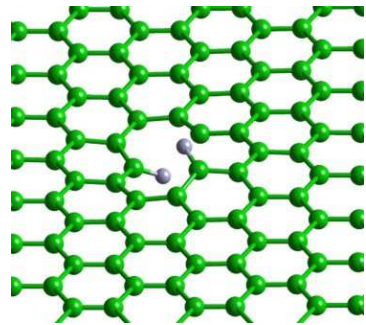}
\caption{The structure of the defect considered in the present study.}
\end{center}
\end{figure}
Within a valence bond picture
 there are two unpaired electrons in the defect as the breaking of three 
$\pi$-bonds generates 1/3 of an unpaired electron on each of the C-atoms 
neighbouring the vacancy and the breaking of the $\sigma$-bond also generates an
 unpaired electron. Hund's rule coupling within the uncoordinated C-atom and 
the spin-alternation rule operating between the  $\pi$-electrons ensures that 
these spins are aligned and thus the overall magnetic moment  on the defect is
 2$\mu_B$ ($\mu_B$ is the Bohr magneton). \footnote{We note that the
   discrepancy with the value of 1.2 $\mu_B$
obtained in Ref.~\cite{8} for the same defect is due to 
a higher degree of electron localisation induced by a hybrid functional.}
In the current study this defect is used as 
an example of the many possible structures at which electron localisation 
might occur in a graphene sheet. The conclusions drawn below depend on the arrangement
of defects and on their ability to generate a local magnetic moment through
an instability of the electronic system, but are not expected to depend
on the structural and chemical details of the defects.

\begin{figure}
\begin{center}
\vspace{.5cm}
\includegraphics[scale=1.1]{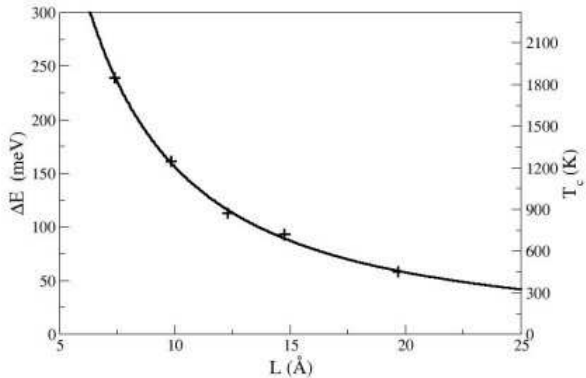}
\caption{The dependance of the magnetic stabilization energy $\Delta$E on the 
distance between defects $L$.}
\end{center}
\end{figure}

The interaction of the localised moments is quantified by computing
the quantity $\Delta E$ (stabilization energy) which
is defined as the difference between the total 
electronic energy of the antialigned and aligned arrangements of the defects' 
spins. The decay of the stablisation energy with $L$ is
approximately $\Delta E=4208 L^{-1.43}$ meV and it is shown in Fig. 2. 
The calculated coupling energy can be related to the Curie temperature
($T_c$), at which the ferromagnetic order is lost due to
thermal fluctuations, through an effective Heisenberg model with 
an interaction $J=\Delta E/2$ per pair of spins. 
According to the Mermin-Wagner theorem~\cite{18}, a two dimensional isotropic Heisenberg
ferromagnet is unstable with respect to thermal fluctuations at any finite temperature and therefore
does not support long range order.
However such effects are seen to reduce significantly
when a small coupling along the third direction is introduced.
Recent Quantum Monte Carlo (QMC) studies on quasi two dimensional Heisenberg
ferromagnets~\cite{schmal,qmc} have shown that the introduction
of an interlayer coupling 0.1\% of that in-plane 
is enough to restore a finite value of $T_c$.
In particular, for a square lattice of spin-$1$ $T_c=JS(S+1)0.44/K_B$
(see Fig. 3 in Ref.~\cite{qmc}) which, compared to a mean-field expression 
$T_c=JS(S+1)z/3K_B$ (z is the coordinaton number), differs by a factor of about $3$.
On the right-hand side of the graph in Fig. 2, the transition temperature
$T_c$ for a square lattice arrangement of defects within the QMC formula above 
is shown for different interdefects distances. 
For this arrangement of the defects, the ferromagnetic state is lower
in energy over the whole range of distances considered and is predicted to be
stable at room temperature up to very large defect separations (Fig. 2).

The present system should be considered as a model system for the study
of more complex and realistic cases such as defective graphite~\cite{5}
or multilayer defective graphene or defective graphene on a substrate where a small
but non zero exchange interaction in the third direction is usually found.
The ferromagnetic long range order at room temperature is
in agreement with recent experiments on bombarded graphite
where a room temperature ferromagnetic ground state is found and the average inter-defect 
distance was determined to be about 1 nm~\cite{interdef}.

Effects due to interchain interaction expected at distances L close to the interchain
fixed distance of 25 \AA\ are seen not to change the picture qualitatively.
To confirm  this, we independently considered a series of calculation of  $\Delta E$ for a
strictly 2-dimensional arrangement of defects (a hexagonal superlattice of defects)
for which the qualitative nature of the long ranged interaction is unchanged and
the power of the decay law is found to be 1.7. 

The bipartite nature of the lattice and the presence of the partially filled 
$\pi$ system of electrons in the $sp^2$ network of atoms are essential for 
establishing the strong, long-range coupling mechanism that gives magnetic 
ordering at high temperature. This mechanism is the result of local exchange 
repulsion, which ensures that the partially filled $\pi$ orbital of each carbon 
atom is coupled in an anti-parallel fashion to that on its nearest neighbour.
The operation of this mechanism to produce a long range spin polarisation 
within the sheet is evident in the spin density displayed in Fig. 3 and has been 
observed previously in $\pi$ bonded carbon networks~\cite{19}.
\vspace{.2cm}
\begin{figure}[!h]
\begin{center}
\includegraphics[scale=.6]{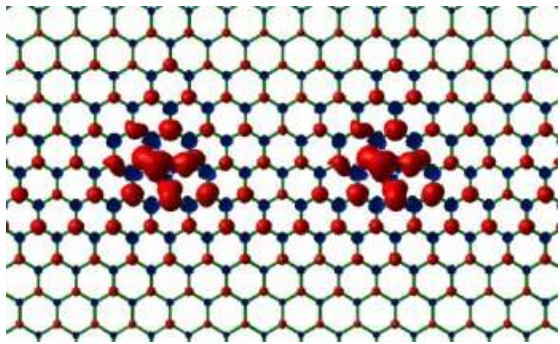}
\caption{(color on line) Isovalue surfaces of the spin density of the graphene sheet with a 
defect separation of 20 \AA . The red (at 0.021 $\mu_B/$\AA) and blue (at -0.021  
$\mu_B/$\AA))  isosurfaces represent the majority and minority spin densities,
 respectively. The majority spin density within the defect is concentrated on the
 unsaturated carbon atoms and their second neighbours. The operation of the 
spin alternation rule in the spin polarised lattice is clearly visible.}
\end{center}
\end{figure}
It is notable that this mechanism extends the validity of Lieb's theorem 
beyond a nearest neighbour model~\cite{20}. The predicted ferromagnetic nature of the 
overall interaction is also entirely consistent with previous studies of 
interactions between extended defects in graphene ribbons~\cite{9,10}. 
It is clear from the spin alternation mechanism that the ferromagnetic 
coupling between the localised spin moments depends on the defects being 
arranged on the same sublattice of the graphene sheet. 
A regular arrangement on opposite sublattices would lead to 
antiferromagnetic coupling, as seen by Yazyev {\it et al.} at short
separation~\cite{yaz07}, and a random arrangement of defects would be
expected to lead to a ground state of smaller spontaneous magnetisation,
while retaining the strong local coupling. 
This prediction agrees with the current experimental 
observation by Esquinazi {\it et. al.}~\cite{5,interdef} (i.e. high
transition temperature and small total magnetisation), 
where the defect density is controlled, to an extent, via
the defect implantation technique but the defects are most probably distributed randomly.
The production of a graphene sheet and control of its defect density is not unrealistic 
using current techniques~\cite{2,5} but methods for controlling 
the defect arrangement have not yet been established.

To analyse the semiconducting properties of the electronic structure 
of the system considered in the present study,
the electronic band structure is show in Fig. 4 for the ferromagnetic state 
of a triangular array of defects at a separation of ~20 \AA\ . 
This choice of array facilitates the comparison with that electronic
structure of perfect graphene. 
The previously analysed structures (parallel chains and hexagonal arrangement of defects)
show the same qualitative features as for the triangular array.

The presence of the defects and the magnetic ordering 
break the symmetry of the graphene $\pi$ system, opening band gaps  
of 0.51 and 0.55 eV in the majority and minority spin band structures, respectively.
The energy gap of the majority states is shifted
upwards by about 0.20 eV with respect to the energy
gap of the minority states, and the shape of the bands around the Fermi energy
is markedly spin dependent. A strong spin asymmetry in the conductivity of the
sheet is therefore to be expected. The size of the band gap is reduced to 0.1 eV
when adopting the PBE generalised gradient approximation~\cite{pbe}.

Calculations at a variety of defect separations establish that the band gap
scales as  $L^{-2}$. This decay law is consistent with that recently
calculated for graphene ribbons where the edges are line defects at which the
spin moments localise and the band gap scales as $L^{-1}$ with the ribbon width.

\vspace{.2cm}
\begin{figure}[!h]
\begin{center}
\includegraphics[scale=1.]{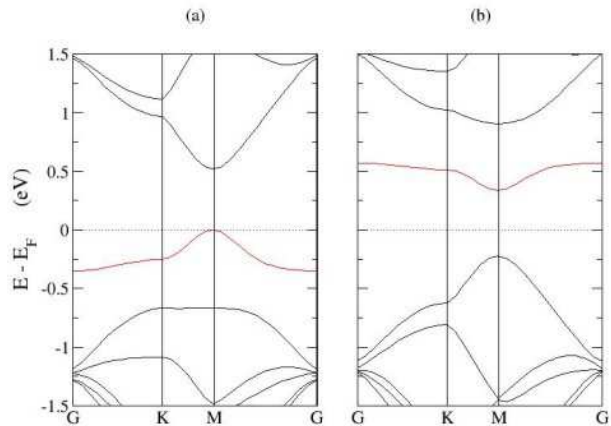}
\caption{(color on line) The electronic energy bands of the majority (a) and minority (b) 
spin states of the ferromagnetic ground state in a defective graphene sheet with 
a defect separation of 20 \AA~plotted with respect to the Fermi energy
(E$_F$).
The rather flat impurity bands near the Fermi energy, which are associated with 
the states localised at the defect, are indicated in red.}
\end{center}
\end{figure}
Recently Yazyev {\em et al.}~\cite{yaz07} reported GGA calculations that produced
ferromagnetism in graphene
induced by the presence of defects but the magnetic coupling mechanism proposed
differs strongly from that described in the current work.
In Ref.~\cite{yaz07} a metallic state is obtained and the ferromagnetic order is 
therefore assumed to be supported by itinerant electrons. 
In the present study the ground state is semiconducting and the magnetic coupling
is clearly not due to itinerant electrons.
For this reason we adopt an effective Heisenberg model
in order to assess the temperature related properties of the system
rather than the Stoner model of Ref.~\cite{yaz07}.
The explanation for this different behaviour does not reside in the morphology
or in the concentration of defects nor in the treatment of electronic exchange and correlation
(the band gap is still present when using the gradient corrected
functional PBE as adopted in Ref.~\cite{yaz07}) but is due to the arrangement
of defects. In Ref.~\cite{lindan}, the authors study the same H-chemisorption defect
as in Ref.~\cite{yaz07}, but consider a different arrangement of defects
(specifically, the unit cell in  Ref.~\cite{lindan} has zig-zag borders whereas 
armchair borders are used in Ref.~\cite{yaz07}). 
This results in a semiconducting ground state as found here.
In Ref.~\cite{lindan} the magnetic coupling between defects
was not analysed and this is the aim of the present study.

The opening of a gap obtained in this study 
is an effect of much larger magnitude than the opening due to spin-orbit coupling
described in Ref.~\cite{dug06}. The effect of the spin-orbit coupling
is not included in our study since it is expected to be negligible in
carbon systems~\cite{tra07}.

\section{Conclusions}
\label{conc}
In conclusion the ground state of a two dimensional graphene sheet containing a 
periodic array of point defects  has been shown to be both 
ferromagnetic at room temperature and semiconducting for defect separations up 
to 20 \AA\ and thus for defect concentrations as 
low as 10$^{13}$/cm$^2$. The energy gap and 
magnetic coupling depend strongly on defect concentration.  
We conclude that a 
doped or defective graphene sheet is a very promising material with an in built 
mechanism for tailoring properties for a variety of spintronics applications.

\section*{Acknowledgments}
This work is supported by the European Union under the NEST FERROCARBON
project (CEC 012881). The authors would like to thank G. Mallia for technical
and scientific  assistance, the Computational Science and Engineering
Department of the STFC for providing the computing facilities, and the
Materials Chemistry Consortium for providing computer time.



\begin{thebibliography}{99}

\bibitem{1} Wolf S A, Awschalom D D, Buhrman R A, Daughton J M, von
  Molnar S, Roukes M L, Chtchelkanova A Y , and Treger D M, 2001 {\it Science} {\bf 294}, 1488

\bibitem{2} Novoselov K S, Geim A K, Morozov S V, Jiang D, Zhang
  Y,Dubonos S V, Grigorieva I V, and Firsov A A, 2004, {\it Science} {\bf 306}, 666

\bibitem{4} {\it Carbon-based Magnetism: an Overview of the Magnetism of
Metal-free Carbon-based Compounds and Materials}, edited by T. Makarova and F. Palacio, Elsevier,
Amsterdam, 2006.

\bibitem{5} Esquinazi P, Spemann D, Hohne R, Setzer A, Han K H, and
  Butz T, 2003, {\it Phys. Rev. Lett.} {\bf 91}, 227201

\bibitem{xmcd} Ohldag H, Tyliszczak T, Hohne R, Spemann D, Esquinazi
  P, Ungureanu M, and Butz T, 2007m {\it Physical Review
  Letters} {\bf 98}, 187204

\bibitem{6} Murata K, Ushijima H, Ueda H, Kawaguchi K, 1992, {\it
  J. Chem. Soc. Chem. Commun.}, {\bf 567}

\bibitem{7} Ref [1] pag. 437-463.

\bibitem{8} Lehtinen P O, Foster A S, Ma Y, Krasheninnikov A V, and
  Nieminen R M, 2004,
{\it Phys. Rev. Lett.} {\bf 93}, 187202

\bibitem{9} Pisani L, Chan J A, Montanari B, Harrison N M, 2007, {\it Phys. Rev. B} {\bf 75}, 064418

\bibitem{10} Son Y -W, Cohen M, Louie S G, 2006 {\it Nature} {\bf 444}, 347

\bibitem{11} Son Y -W, Cohen M, Louie S G, 2006, {\it Phys. Rev. Lett.} {\bf 97}, 216803

\bibitem{yaz07} Yazyev O V and L. Helm, 2007, {\it Phys. Rev. B} {\bf 75},
  125408

\bibitem{voz} Vozmediano M A H, Lopez-Sancho M P, Stauber T and Guinea
  F, 2005, {\it Phys. Rev. B} {\bf  72}, 155121

\bibitem{sar} Saremi S 2005 {\it Phys. Rev. B} {\bf  76} 184430

\bibitem{dug06} Dugaev V K, Livinov V I, and Barnas J, 2006
  {\it Phys. Rev. B} {\bf 74}, 224438

\bibitem{edw} Edwards D M and Katsnelson M I, 2006 {\it J.Phys.:Cond.Mat.} {\bf 18} , 7209 

\bibitem{12} Becke A D, 1988, {\it Phys. Rev. A} {\bf 38}, 3098

\bibitem{13} Becke A D, 1993, {\it J. Chem. Phys.} {\bf 98}, 5648

\bibitem{14} Lee C, Yang W, and Parr R G, 1988, {\it Phys. Rev. B} {\bf 37}, 785

\bibitem{15} Muscat J, Wander A, Harrison N M, 2001, {\it Chem. Phys. Lett.} {\bf 342}, 397

\bibitem{16} Martin R L, Illas F, 1997, {\it Phys. Rev. Lett.} {\bf 79}, 1539

\bibitem{17} Saunders V R, Dovesi R, Roetti C, Orlando R, 
Zicovich-Wilson C M, Harrison N M, Doll K, Civalleri B, Bush I J,
D'Arco Ph, and Llunell M, 2003 {\it CRYSTAL2003 User's Manual},
(University of Torino, Torino)

\bibitem{17a} Gale J D, 2003, {\it General Utility Lattice Program},
 (http://gulp.curtin.edu.au/)

\bibitem{18} Mermin N D, Wagner H, 1966, {\it Phys. Rev. Lett.} {\bf 17}, 1133-1136

\bibitem{schmal} Schmalfuss D, Richter J, and Ihle D, 2005 {\it Phys. Rev. B} {\bf 72}, 224405

\bibitem{qmc} Yasuda C {\em et al.}, 2005, {\it Phys. Rev. Lett.} {\bf 94}, 217201

\bibitem{interdef} Barzola-Quiquia J, Esquinazi P, Rothermel M, Spemann D, Butz t, 
and  Garcia N,  2007 {\it Phys. Rev. B} {\bf 76}, 161403

\bibitem{19} Chan J A, Montanari B, Chan W -L, Harrison N M, 2005,
  {\it Mol. Physics} {\bf 310}, 2573-2585

\bibitem{20} Lieb E H, 1989, {\it Phys. Rev. Lett.} {\bf 62}, 1201

\bibitem{lindan} Duplock E J, Scheffler M, and Lindan P J, 2004,
{\it Phys. Rev. Lett.} {\bf 92}, 225502

\bibitem{guinea} Pereira V M {\em et al.}, 2006, {\it Phys. Rev. Lett.} {\bf 96}, 036801

\bibitem{tra07} See for instance Trauzettel B, Bulaev D V, Loss D, and
  Burkard G, 2007, {\it Nature Physics} {\bf 3}, 192

\bibitem{pbe} Perdew J P, Burke K, and Ernzerhof M, 1996
{\it Phys. Rev. Lett.} {\bf 77}, 3865

\end{thebibliography}
\end{document}